# PREDICTIVE NETWORKING AND OPTIMIZATION FOR FLOW-BASED NETWORKS.

by

**MICHAEL ARNOLD**

**A THESIS**

In presenting this thesis in partial fulfillment of the requirements for a master's degree from The University of Alabama in Huntsville, I agree that the Library of this University shall make it freely available for inspection. I further agree that permission for extensive copying for scholarly purposes may be granted by my advisor or, in his/her absence, by the Chair of the Department or the Dean of the School of Graduate Studies. It is also understood that due recognition shall be given to me and to The University of Alabama in Huntsville in any scholarly use which may be made of any material in this thesis.

Michael Arnold (date)

## THESIS APPROVAL FORM

Submitted by Michael Arnold in partial fulfillment of the requirements for the degree of Master of Science in Computer Science and accepted on behalf of the Faculty of the School of Graduate Studies by the thesis committee.

We, the undersigned members of the Graduate Faculty of The University of Alabama in Huntsville, certify that we have advised and/or supervised the candidate of the work described in this thesis. We further certify that we have reviewed the thesis manuscript and approve it in partial fulfillment of the requirements for the degree of Master of Science in Computer Science.

| Signature | Date | Role |
|---|---|---|
| *Dr. Daniel Rochowiak* | (Date) | Committee Chair |
| *Dr. Harry Delugach* | (Date) | |
| *Dr. Letha Etzkorn* | (Date) | |
| *Dr. Heggere Ranganath* | (Date) | Department Chair |
| *Dr. Sundar Christopher* | (Date) | College Dean |
| *Dr. David Berkowitz* | (Date) | Graduate Dean |

## ABSTRACT

School of Graduate Studies
The University of Alabama in Huntsville

Degree Masters of Science in Computer Science College/Dept. College of Science/Computer Science

Name of Candidate Michael Arnold

Title Predictive Networking and Optimization For Flow Based Networks

Artificial Neural Networks (ANNs) were used to classify neural network flows by flow size. After training the neural network was able to predict the size of a flows with 87% accuracy with a Feed Forward Neural Network. This demonstrates that flow based routers can prioritize candidate flows with a predicted large number of packets for priority insertion into hardware content-addressable memory.

Abstract Approval: Committee Chair ______________________
*Dr. Daniel Rochowiak*

Department Chair ______________________
*Dr. Heggere Ranganath*

Graduate Dean ______________________
*Dr. David Berkowitz*

## ACKNOWLEDGMENTS

When I first started working on this thesis, I had this silly idea that I would have the work done in no time and that I could do it without any support from anyone. Reality is a harsh professor, and the reality is that I received a lot of help and support from numerous persons, friends, and family.

First, my family was instrumental in the success of this thesis; the regular "is your thesis done yet" gave me the motivation I needed to continue. I also need to thank my wife, Jia Chen, as she was extremely patient and supportive during the long hours required to complete this thesis. My grandfather deserves special mention. He is the reason why I became an engineer in the first place. I will never forget those summers spent tinkering with various machinations in your garage. You taught me how to solve large problems. "It is like eating an elephant" you said. You take it "one bite at a time".

My long time friend, Dr. Kornman, helped me in more ways than he will likely ever know. His influence was very early in my academic life, but it should not be discounted. I would also like to mention JP Viljoen. Your insights are always helpful, and always welcome. Last, but not least, I would like to thank Dr. Rochowiak, Dr. Etzkorn, and Dr. Delugach for all the work you put in, and for all of your advice

## TABLE OF CONTENTS

## LIST OF FIGURES

FIGURE PAGE

## LIST OF TABLES

TABLE PAGE

# List of Terms

**CAM** Content Addressable memory.

**CPE** Customer Premise Equipment.

**Network Function Virtualization** Taking network functionality that typically runs on dedicated hardware, such as a router, and placing it in a virtual machine.

**Network Virtualization** Using technologies such as VRFs and VLANs in order to achieve tenant isolation.

**Open Network Foundation** An organization dedicated to accelerating the adoption of SDN and NFV.

**Software Defined Networking** Separation of network's control-plane from the network's data-plane; the control functionality is separate from the forwarding functionality.

**Virtual Routing and Forwarding** Isolation of routing functionality and IP functionality.

*This is dedicated to my grandfather. Thank you for fueling my interest in science and engineering.*

# CHAPTER 1

# INTRODUCTION

## 1.1 Introduction

### 1.1.1 Brief History of Software Defined Networking

Software Defined Networking (SDN) was created by the Open Network Foundation (ONF) in an effort to remedy several complaints by a broad range of network operators that current network architectures are not suitable to meet the requirements of today's enterprises. More precisely, the ONF was looking for a network architecture that was centralized, had a well defined API so it could be user extensible / programmable, could scale well, and break vendor dependence. [3]. To justify this need, the ONF cited changing traffic patterns, consumerization of IT, the success of the cloud abstraction of services, and the bandwidth requirements of current applications as the driving factors. [3]. The SDN model is an attempt to change this by separating the control-plane from the data-plane with commoditized hardware on the data-plane, and a centralized SDN Controller on the control-plane. [3]

There are three layers to SDN. The application plane is a set of applications that sit on top of the SDN Northbound Interface and programmatically communi-cate the desired network behavior to the SDN controller. [4] These SDN applications can be virtual instances of routers, load balancers, L2 switches, intrusion detection systems, even load balancing at the flow level. The Northbound Interfaces are the APIs which applications utilize to communicate with the SDN Controller; the SDN Controller is responsible for translating between the application plane and the data plane and houses the control logic. The SDN data-plane is where the forwarding engine manipulates content-addressable memory (CAMs) on the physical hardware.

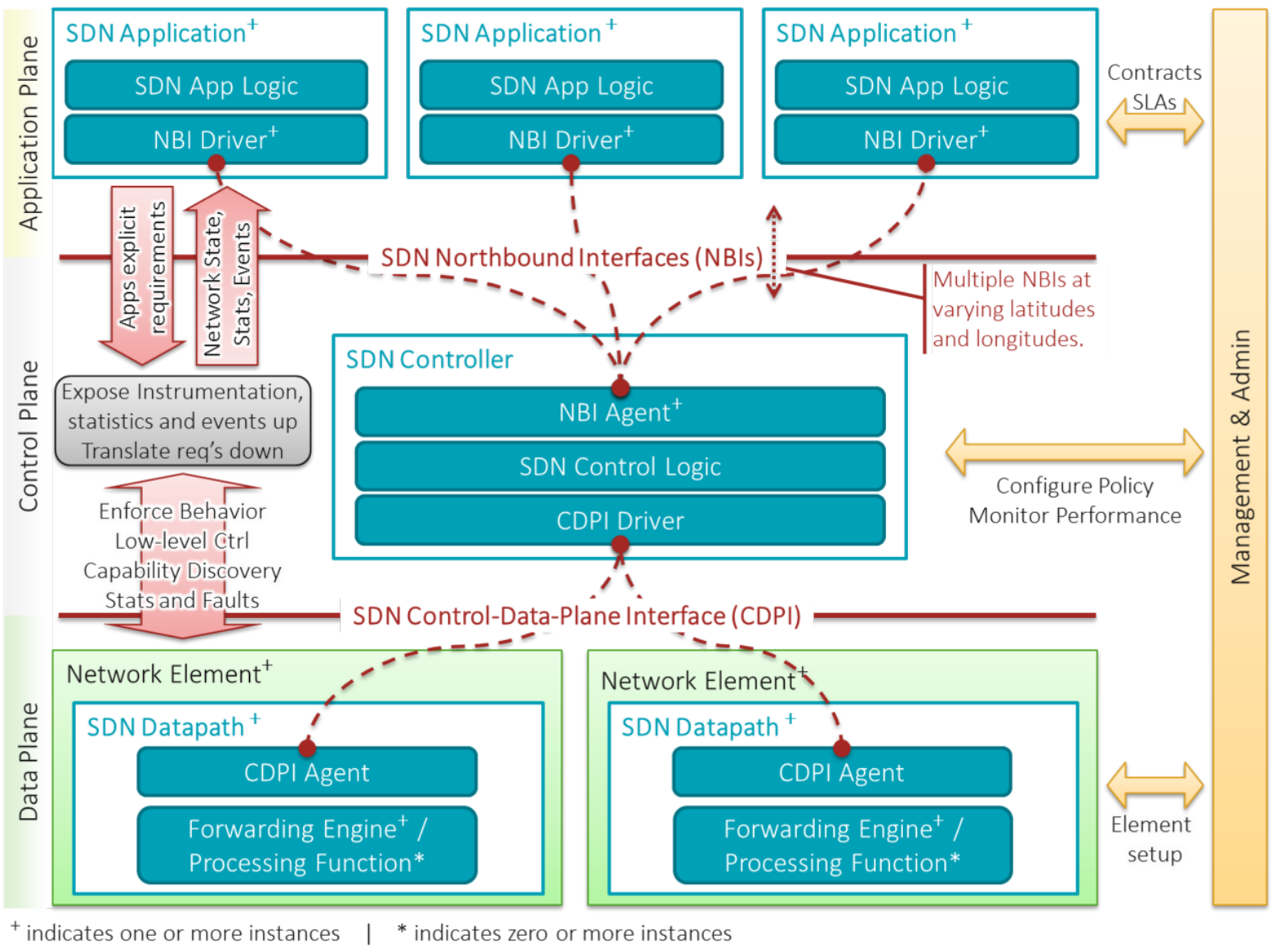

**Figure 1.1**: SDN Architecture. [1]

Initially, Network Function Virtualization (NFV) is going to be the biggest use case for [5] SDN. An example of networking function virtualization is the use of virtualized router components such as a virtual machine. The reason for this is so network operators can more efficiently use resources and create a realization of Infrastructure as a Service. Long term, the data-plane control-plane split abstraction is going to be the key benefit to SDN; it allows for a never-before-seen level of interoperability, programmability, and extensibility. For example in the CORD (Central Office Re-architected as a Datacenter) model [6] an OpenFlow top of rack (TOR) switch is used as a dataplane contolled by the OpenFlow controller. The TOR switch is also used to steer traffic to network functions which are run on virtual machines. This allows for the hardware to become a commidty while allowing any number of vendors to provide network functions such as SIP phones, vRouters, firewalls, proxies, or even the control plane for customer premises equipment (CPE) through the use of cloud extension devices. NFV is not to be confused with network virtualization which is technologies such as VRFs, VLANs, VXLAN. In this context these technologies would be used achieve tenant isolation. [7].

### 1.1.2 Neural networks applied to computer networking

Neural networks have been shown to hold significant improvements in computer networking, such as seen in [8] where the author demonstrates a novel approach to shortest-path-tree computation using pulse-coupled neural networks which were previously used and invented for use in the image segmentation problem. Pulse-coupled neural networks are a class of biologically inspired neural networks modeled on a cat's visual cortex [8]. Another example is [9] where the author demonstrates using Hopfield neural networks for a shortest path routing algorithm. Hopfield neural networks are a form of RNN (Recurrent Neural Network) that are guaranteed to converge to a local minimum, allowing them to be used to serve as CAMs (Content Addressable Memory) which would be used to provide fast hardware based lookup ta-bles. The question, however, remains. Can we use neural networks to solve problems that were not previously solvable in computer networking?

### 1.1.3 Purpose of the Study

The literature has shown that neural networks have the capability of matching complex patterns, including those over time. The literature has also shown that complex actions can be taken as a result of pattern recognition on neural networks, but several questions remain which are answered in Chapter 5.

1. Can a neural network match patterns in flow data in real time and be used to optimize the use of resources?

2. Can a neural network predict future resource utilization based on current inputs?

3. Can these techniques be applied to dynamically and preemptively route around predicted points of congestion?

In summary, neural networks have the ability to profoundly change the future of computer networking, especially with the advent of SDN. This study will attempt to demonstrate the power of neural networks in making decisions in flow based systems. The next section will build the ground work for defining the conditions needed, including a definition for OpenFlow, an introduction to machine learning, and different types of neural networks.

# CHAPTER 2

# BACKGROUND

## 2.1 OpenFlow Protocol

"An OpenFlow switch is a software switch which consists of one or more pipelined flow tables, a group table, which performs packet lookups and forwarding, and an OpenFlow channel to an external controller." [2]. The flow table is essentially a lookup table with match fields and actions, and is processed like a pipeline. Pipelined flow tables contain traffic flows, as defined later. A flow will match the first flow table, and potentially be forwarded to a port or another flow table. This is what we mean by pipelined; the flow match rules happen iteratively, like water flowing through a pipe. A traffic flow is a "sequence of packets sent from a particular source to a particular unicast, anycast, or multicast destination that the source desires to label as a flow" [10]. Flow classifiers are typically based on the 5-tuple consisting of destination address, source address, protocol, destination port, source port. The primary benefit of flow based routing is that it eliminates the need to do lookups to the routing table on a per-packet basis. The route lookup can be done for the first packet in a flow, and then the same transform applied to each packet in the sequence. Flow tables can easily be implemented in hardware, and most vendors support some form of flow matching in either software or hardware ternary content-addressable memory (TCAMs). [11]

In the SDN model, OpenFlow serves as the data plane handling packet forwarding operations for the OpenFlow controller [2]. The flow tables handle packet lookups and forwarding. The flow table contains match fields, priority bits, counters, instructions, timeouts, cookies, and other things used to match a flow via the 5-tuple. Once the flow is matched, an action is then applied such as pushing and popping tags, setting header fields, TTL decrements, among others. Think of this as an if then rule. If the frame matches this 5-tuple, then we apply this action set. For example, consider an example for an L3 router. An L3 router is a router which performs forwarding decisions based on the L3 Internet Protocol (IP) payload. An L3 packet comes in and is sent to the ingress flow table, which is matched by the table-miss flow entry. This flow entry will then forward the packet to the controller for a route lookup. The controller finds the appropriate next hop and the proper network interface, and pushes a new flow entry to the OpenFlow switch for this packet and forwards it out the appropriate interface. The next packet in that flow will match the flow entry that was just pushed down into the OpenFlow switch, which will then apply the action to the packet forwarding it out the same egress interface the previous packet was sent to, and applying the same action. Only the first packet in a flow would cause a route lookup, speeding up packet processing. [11]

### 2.1.1 OpenFlow Switch Components

An OpenFlow switch consists of the following components as shown in Figure 2.1: OpenFlow Channel, Group Table, and one or more Flow Tables. The switch communicates with the controller via the OpenFlow Channel, which is the communications protocol used. Using the OpenFlow Channel, the controller can add, update and delete flow entries in flow tables both reactively and proactively. Each flow entry consists of various match fields, counters, and instructions. [2]

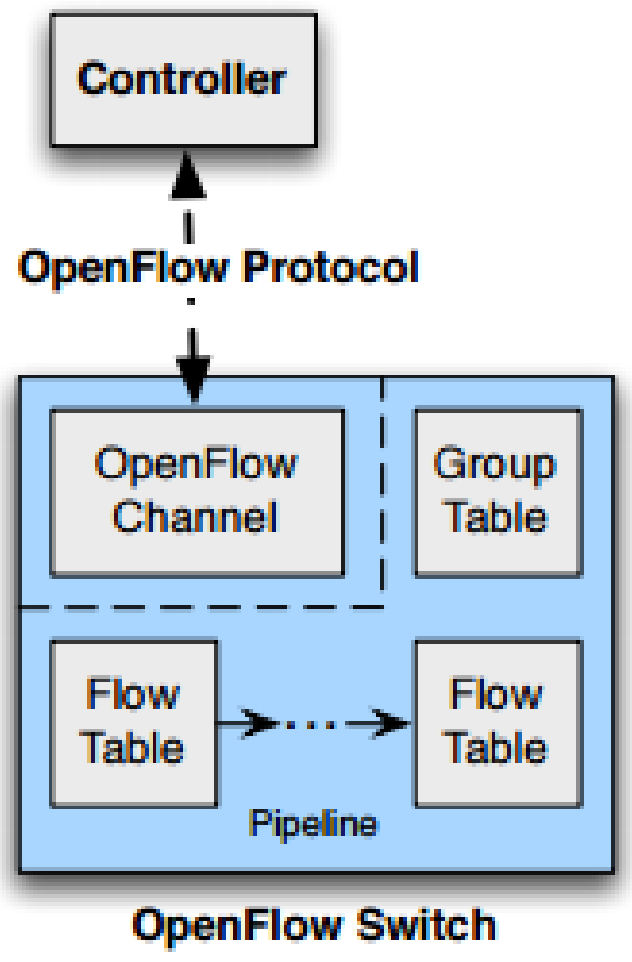

**Figure 2.1**: OpenFlow Switch Components [2]

### 2.1.2 OpenFlow Ports

OpenFlow ports are logical network interfaces for use in passing packets between OpenFlow Processing units and the rest of the network. OpenFlow switches also connect to the controller, and other switches via OpenFlow ports. [2]

#### 2.1.2.1 Standard Ports

Standard Ports are physical ports, logical ports, and the LOCAL reserved port exclusive of all other reserved ports. These are the only ports which can be used in groups, and that have port counters. [2]

#### 2.1.2.2 Physical Ports

Physical ports are switch defined ports which correspond to a hardware interface on the switch. The term 'Physical' is used rather loosely here, as these can be virtualized interfaces, such as in SR-IOV (Single Root IO Virtualization) which carves a single physical interface into virtual slices. [2]

#### 2.1.2.3 Logical Ports

An OpenFlow logical port is a switch defined port that correlates with hardware interfaces; these are abstractions on physical interfaces. For instance, these could be tunnel interfaces, link aggregation groups, loopback interfaces, among others. [2]

#### 2.1.2.4 Reserved Ports

Reserved Ports represent forwarding actions. [2]. See table Table 2.1 for the various types of forwarding actions and their definitions. [2] Recall that a flow router consists of a lookup table and an action set; the following definitions define the available action sets.

| | |
|---|---|
| ALL: | Copies the packet to all standard ports |
| CONTROLLER: | Wraps the frame in a PACKET_IN message and sends the frame to the controller. If this is used as an ingress port, then this message would contain the packets originating from the controller. This is used to create rules that forward the complete frame to the controller for further processing, such as when you do not yet have an entry in the flow table to match this type of packet; think default rule |
| TABLE: | This is the start of the OpenFlow pipeline. It is used as an output action and drops the packet in the first table in the table pipeline for processing. |
| IN_PORT: | Represents the packets input port; this will send the packet back from where it came (hairpinning). |
| ANY: | Represents a wildcard value in a field. |
| LOCAL: | Represents the switches local networking and management stacks. |
| NORMAL: | Represents the non-OpenFlow processing pipeline. |
| FLOOD: | The frame will be sent to all output ports, but not the ingress port. |

**Table 2.1**: OpenFlow Forwarding Actions

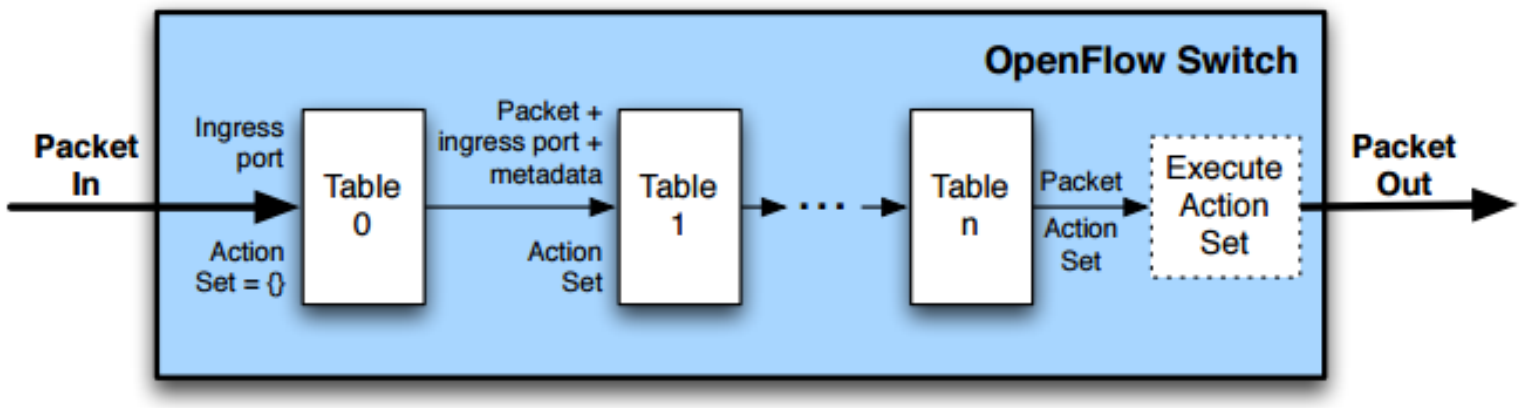

**Figure 2.2**: Flow Table Packet Processing Pipeline [2]

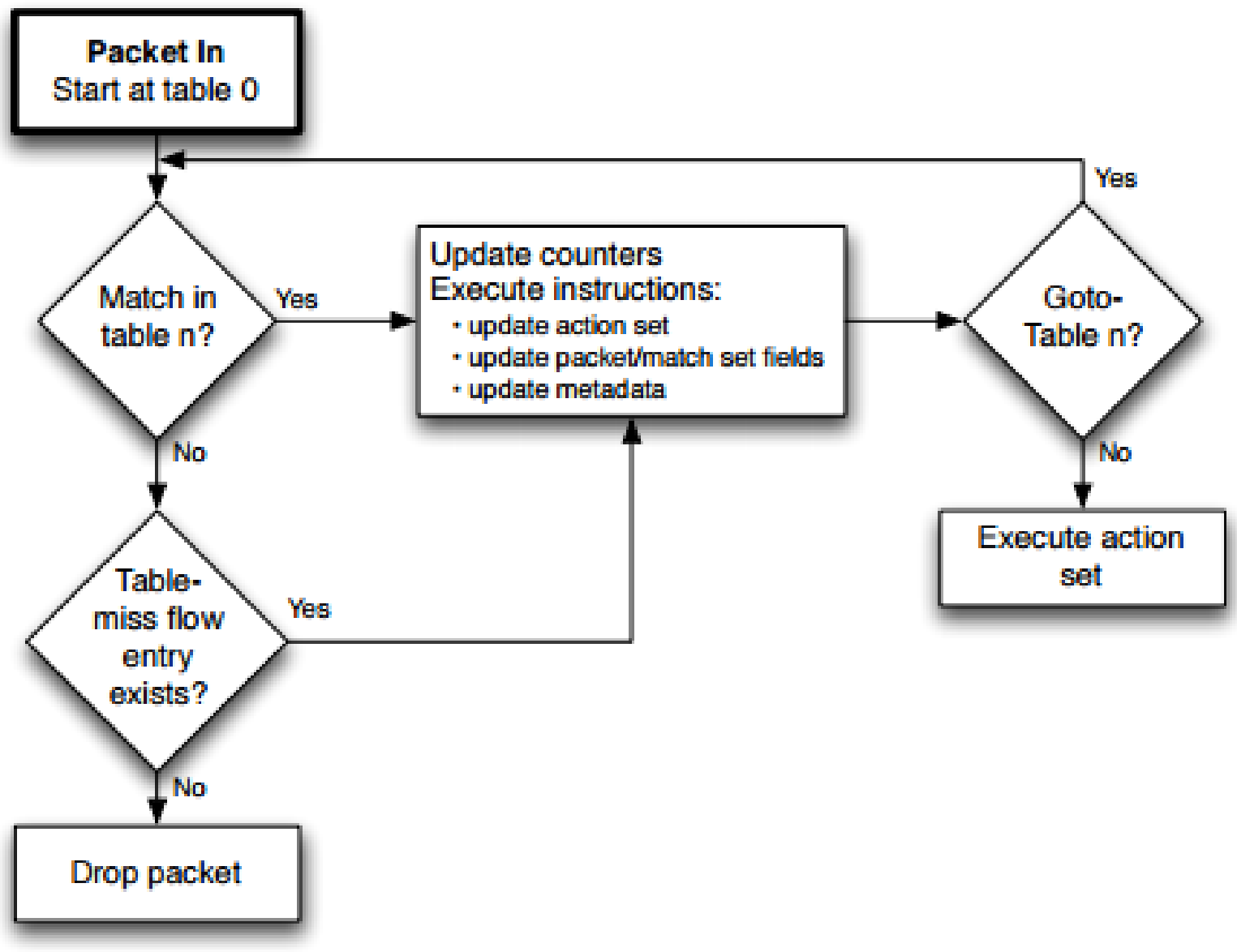

**Figure 2.3**: Flow Matching Flow Chart [2]

### 2.1.3 OpenFlow Flow Table

The flow table is essentially a lookup table with match fields and actions, and is processed like a pipeline. See Figure 2.2 and Figure 2.3. When the frame ingresses the port it is processed by Table 0 by the highest-priority matching flow entry. This flow entry will contain an action set which can either output the frame to a specific port, apply actions, or send the frame to another table. In the event of a table miss the frame is dropped by the switch. A table miss happens when there is no match rule in the table to match the frame. Each Flow Table contains the following fields [2]. Recall that a flow router consists of a lookup table and an action set; this is the lookup table that matches on various fields in the packet header.

| | |
|---|---|
| Match Fields: | The match criteria for frames. Consists of header data and metadata information. Match fields are placed on the flow table in order to define the packet to which an action is to be performed. This contains the 5-tuple information and some additional criteria that can also be used. |
| Priority: | The match priority. Matches occur in priority order (the highest priority match wins). Useful for defining exception entries and default entries in the table pipeline. |
| Counters: | Counts the number of matches. |
| Instructions: | Defines what is to be done to the frame after a match; there are one or more of these. |
| Timeouts: | Defines how long a flow can exist in the switch. A soft timeout defines how long the flow lives if a matching frame has not been seen. A hard timeout defines how long a flow lives no matter the match count. |
| Cookie: | Controller defined field. This is not used in packet processing but is useful for filtering flow statistics. |

**Table 2.2**: OpenFlow Flow Table Fields

#### 2.1.3.1 Flow Removal and Eviction

Flows can be removed from the controller in three ways: at the request of the controller, by expiration, or via the switch eviction mechanism. [2] The Flow switch expiration mechanism defines a hard and an idle timeout. The idle-timeout causes eviction if and only if no frames have been seen for the duration. The hard-timeout causes eviction no matter if matches have been seen. The controller can also send a DELETE message, causing flow removal. The flow switch eviction mechanism lets the switch evict flows in order to reclaim resources. Upon removal, a FLOW_REM message may be optionally sent if the SEND_FLOW_REM flag is set in the flow entry. [2] This message is used to inform the controller that a flow has been evicted so it can either keep statistics, or make decisions based on this information.

### 2.1.4 OpenFlow Messages Types

There are three general types of OpenFlow message: controller-to-switch, asynchronous, and symmetric. The controller-to-switch messages are used by the controller to query information from, transmit packets to, or configure the switch. Asynchronous messages are sent without solicitation from the controller. Examples of these include packet-in messages, flow-removed, port-status, and packet-out messages. Symmetric messages require a response from the receiving party. Examples of these are hello, echo, error, and experimenter messages. [2]

OpenFlow defines a specification that one can use to talk to OpenFlow switches, and this technology will be utilized throughout this thesis to provide the mechanism to insert flows into these switches.

## 2.2 Machine Learning

In this section we will provide a taxonomy of learning as it relates to the field of machine learning in artificial intelligence. Throughout this section, we will cite examples of each type of learning, the categories of learning problems, and the types of reasoning which can be used to solve learning problems.

Machine learning is generally broken down into three cases, supervised, unsupervised, and reinforcement learning [12]. Supervised learning is learning by examples of what is good, and what is bad. For instance, consider the case where we are trying to build an autonomous driver for a vehicle. If we are to drive the vehicle while the agent watches, this is supervised learning; we are providing examples of good driving for it to learn from. On the other end of the spectrum is unsupervised learning. This form of learning does not provide any feedback or examples from which the agent to learn; an example of this is clustering in bioinformatics. The agent is not provided examples of the type of gene you are looking for and had no feedback as to how good the match is; the agent simply uses clustering and alignment algorithms to try to find patterns in the data leaving validation to the user [13]. Finally, there is reinforcement learning. This form of learning is similar to supervised learning, except instead of a set of examples being provided as feedback it simply observes the results of its actions. For instance, a chess agent could learn by simply seeing the "win" state at the conclusion of the chess match.

Problems in machine learning can also be broken down into two broad categories, classification problems, and regression problems [12, 13]. If a problem is discrete, it is a classification problem, for example, you may be classifying potential credit card customer by risk, or you could be classifying a set of ailments by which diagnosis they most accurately fit [13]. If a problem is continuous, it is a regression problem, for example, consider the case where you need to fit a series of points to a linear equation [13]. Any example where you are trying to make continuous predictions is a regression problem.

There are three types of reasoning which can be utilized in machine learning: induction, deduction, and abduction [12,13]. Induction entails deriving a model from a set of data; the model, in this case, is the hypothesis that explains the data. Ac-cording to Russel and Norvig [12], "given a collection of examples `f`, return a function `h` that approximates `f`." In this example, h is the hypothesis which approximates `f`. Essentially, inductive learning moves from the specific to the more general, while deduction moves from the more general, to the specific. Abduction, on the other hand, denotes a possible hypothesis, but it does not directly follow from the data; it is a guess based on experience.

### 2.2.1 Supervised Learning

One common example of supervised learning is in training artificial neural networks. According to Haykin [14] a "neural network is a massively parallel distributed processor made up of simple processing units, which has a natural propensity for storing experiential knowledge and making it available for use". Further, the author mentions that artificial neural networks resemble the brain; the network, via machine learning, assimilates knowledge that is stored via the synaptic weights of the pathways.

Artificial neural networks are essentially weighted graphs which provide an input to output mapping as shown in Figure 2.4. Consider a node with three inputs $w_0, w_1, w_2$ such that $w_0$ represents the activation threshold, and $w_1$ and $w_2$ represent the inputs of the node. If $w_1 + w_2 >= w_0$ then the node activates, and we can learn to predict outputs based on inputs if given an example set to learn from.

One technique to learn the activation threshold is called Linear Regression [15] which uses the least-squares cost function to derive the ordinary least squares regression model. Andrew Ng in [15] uses gradient descent to update the edge weights as shown in (2.4), assuming you only have one training example. $\mathtt{h}(\mathtt{x})$ is the hypothesis function.

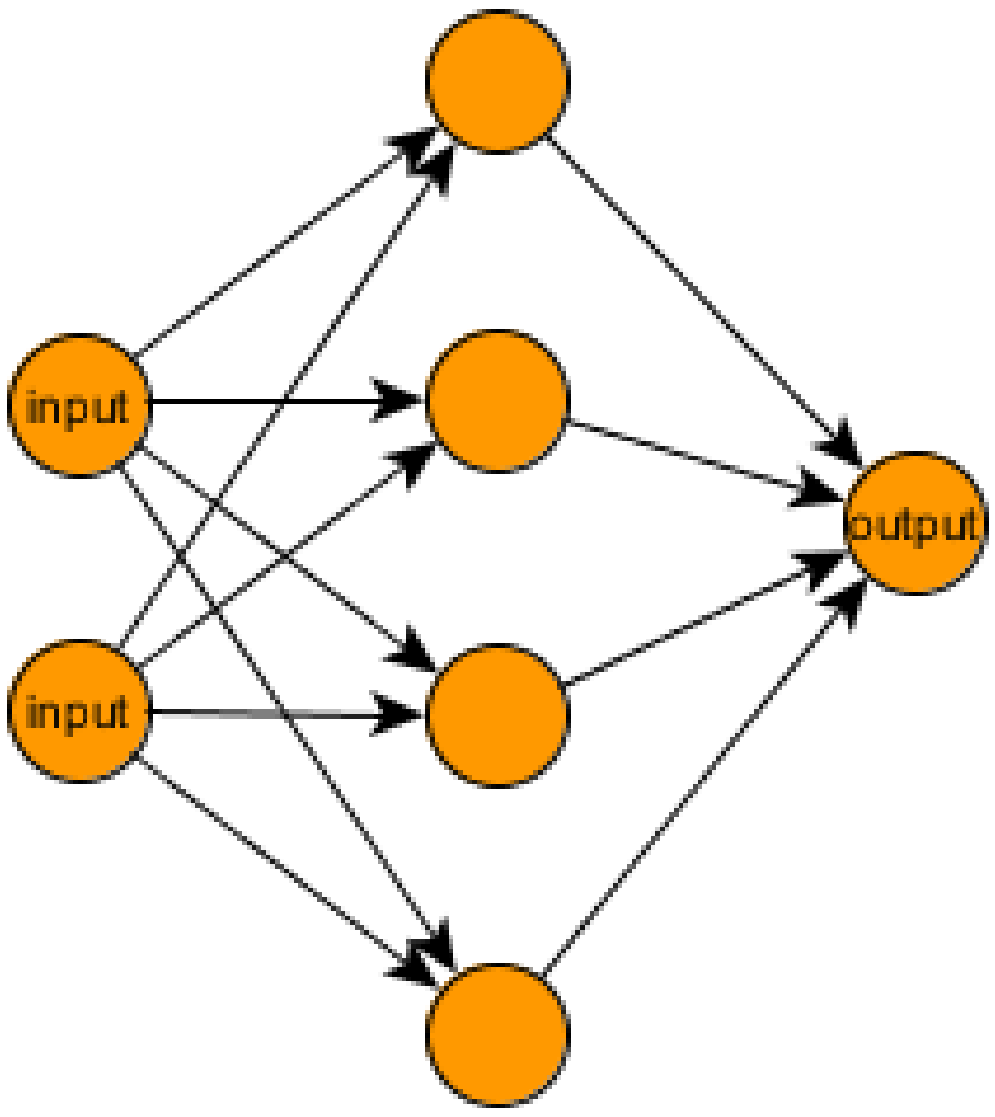

**Figure 2.4**: The graph structure of a neural network.

The derivation is as follows.

$$\theta_j := \theta_j - \alpha \frac{\partial}{\partial \theta_j} J(\theta) \tag{2.1}$$

$$= 2\frac{1}{2}(h_\theta(x) - y)\frac{\partial}{\partial \theta_j}(h_\theta(x) - y) \tag{2.2}$$

$$= (h_\theta(x) - y)\frac{\partial}{\partial \theta_j}(\sum_{i=0}^{n} \theta_i x_i - y) \tag{2.3}$$

$$= (h_\theta(x) - y)x_j \tag{2.4}$$

where $\theta$ is our parameter vector and J is the function to be minimized, $\mathtt{h}$ is the hypothesis, $\mathtt{y}$ is the training example, $\alpha$ is the learning rate, and x is the predicted output. This leads to $\theta_{\mathtt{j}} := \theta_{\mathtt{j}} + \alpha(\mathtt{y} - \mathtt{h}_{\theta}(\mathtt{x}))\mathtt{x}_{\mathtt{j}}$ which is the Least mean squares update rule. [15] The parameter $\alpha$ is our learning rate which when set to zero leads to an inability to learn. The algorithm works by directing the activation threshold in the direction of the provided example. The updates are propagated backwards through the network from the output nodes (back propagation).

### 2.2.2 Unsupervised Learning

Unsupervised learning problems, because of the lack of feedback, often involve data clustering, such as k-means clustering. With clustering algorithms, the idea is to partition the data into k disjoint sets [16]. In general, these types of unsupervised learning algorithms try to find patterns in the data. There are also examples of competitive learning where learning is motivated by competition such as the competition between individuals, or scientists. In neural networks, output neurons compete among themselves to become active. According to Haykin [14], there are three basic elements to a competitive learning rule. First, you need a set of identical neurons with randomly distributed synaptic weights such that they respond differently for different input patterns; higher weighted edges have more influence. Next you need a bias, which controls the strength of the output of the node. Finally, you need a selection mechanism enabling neurons to compete for the right to respond; only one output neuron can respond. It is worth noting that this form of learning allows neural networks to perform clustering, but the idea of competitive learning could have some potential uses in agent based systems as well, allowing agents to compete for a resource in order to determine rules for knowledge discovery, for example.

### 2.2.3 Reinforcement Learning

Reinforcement learning problem solvers generally only learn after the task has been performed. For instance, in a game of chess, the system can learn when it is victorious, however in other environments the rewards can come more frequently [12]. The important thing is that there is some form of reward for good behavior, and a penalty for bad behavior. Reinforcement learning has a lot in common with human learning through trial and error. For instance, consider a player trying to learn to play a strategy game. In cases such as these, simply providing a list of good moves does not work well; there is no way you can possibly learn all of the nuances of a game like chess from a rote list of moves., You can, however, examine the outcomes; if you win, there was something good about what you did, and if you lose then you did something wrong. Because of this, agent based approaches, often times, utilize reinforcement learning rather then supervised learning as it is convenient to simply give the agent the ability to perceive the effects its actions have on the environment as either good or bad. The central theme with this type of learning is exploration; the agent must continuously explore the field in order to experience as much of the environment as possible.

There are several elements of reinforcement learning such as "a policy, reward function, value function, and an optional model of the environment" [17]. The policy is a mapping from the percepts to the actions from which the agent will act. The reward function helps define the goal in the form of payoffs and penalties for good or bad actions, which gives the agent some way to determine how it should alter the policy. Following Sutton et al [17], the reward function is greedy, and does not specify what is good over the long term, rather it shows what is immediately good or bad; the authors liken it to pleasure and pain. The value function, however, defines the agent's long-term goals. Without this function, the agent would ignore immediately bad actions that may payoff in the future with a high value.

Imagine an agent which has a number of states $\mathtt{s}$, and a number of actions $\mathtt{a}$. Let $\mathtt{Q(s,a)}$ be the expected discounted reinforcement of taking action a in state s and $\mathtt{V(s)}$ be the value of s assuming the best action is taken initially. According to Kaebling et al [18] Q learning is defined as $\mathtt{Q(s,a)} := \mathtt{Q(s,a)} + \alpha(\mathtt{r} + \gamma\mathtt{maxQ(s',a')} - \mathtt{Q(s,a)})$ where $\gamma$ is the discount factor, $\mathtt{r}$ is the reward and $\mathtt{maxQ(s',a')}$ is the max future value; this update function takes into account a learning rate $\alpha$ and a discount factor to determine the importance of future rewards.

## 2.3 Neural Networks

### 2.3.1 Structure

A neural network is a network structure modeled after the human brain that is highly complex, nonlinear, and parallel, which has the capability to self organize and adapt in order to solve certain problems and certain computations. [19] In general, there are two basic types of neural networks: Feed Forward Neural Networks (FFNN) and Recurrent Neural Networks (RNNs). A feed forward network is a network such that the input nodes always flow towards the outputs, as opposed to an RNN. In an RNN, there can be cycles such that the output eventually feeds back towards the input. RNNs allow for very complex pattern matching and learning.

Neural networks provide an input-output mapping that is adaptive over time; the synaptic weights change in order to model the behavior of the system. This gives them the ability to generalize, providing answers to questions that it has never seen before. Breaking this taxonomy down further, there are two types of each of these networks: Single and multilayer networks. A single layer network is composed of one or more input neurons, and one or more output neurons. A multi-layer network has one or more input neurons connected to a hidden layer. The hidden layer is then further connected to one or more output neurons. [19]

A neuron is described by the following equations, where k is the neuron, $x_1, x_2, ..., x_m$ are input signals, and $w_{k1}, w_{k2}, ..., w_{km}$ are the synaptic weights [19] and $u_k$ is the linear combiner output due to the input signals; $b_k$ is the bias; $\varphi()$ is the activation function; and $u_k$ is the output signal of the neuron [19] as shown in the following equations. [19]

$$u_k = \sum_{j=1}^{m} w_{kj} x_j \tag{2.5}$$

and

$$u_k = \varphi(u_k + b_k) \tag{2.6}$$

There are three basic types of activation functions [19].

1. Threshold Function:

$$f(x) = \begin{cases} 1, \text{ if} & \mathtt{v} \geq \mathtt{0} \\ \mathtt{0}, & \text{otherwise} \end{cases}$$

2. Piecewise-Linear Function

$$f(x) = \begin{cases} 1, \text{ if} & \mathtt{v} \geq +\frac{1}{2} \\ \text{v, if} & \frac{1}{2} > \mathtt{v} > -\frac{1}{2} \\ \mathtt{0}, & \text{otherwise} \end{cases}$$

3. Sigmoid Function:

$$\varphi(v) = \frac{1}{1 + e^{-av}} \tag{2.7}$$

A neural network can be expressed as a directed graph with four properties:

1. Each neuron is a set of synaptic links, externally applied bias, and an activation function.
2. The synaptic links weight the input signals.
3. The weighted sum of the inputs defines the stimulus neuron.
4. The activation function squashes the stimulus of the neuron. [19]. The process of learning in a neural network is simply the process of updating the weights to model the described output.

### 2.3.2 Forward Activation in a Feed Forward Network

Let the inputs to the neural network be $\hat{\mathtt{i}}$ and the edge weights be $\hat{\mathtt{w}}$ and the activation function be $\mathtt{f(x)}$. Forward activation is computed by taking $\mathtt{f}(\hat{\mathtt{i}}\cdot\hat{\mathtt{w}})$ Typically the activation function is a differentiable function such as the sigmoid function, but it can also be a step function.

### 2.3.3 Back-Propagation in a Feed Forward Network

Back-propagation is the most common method for supervised training on multi-layer neural networks [20] and is typically implemented along with gradient-descent [21]. The first step of back-propagation is to feed the training pattern to the network and generate output activations. [12] Once we have the output activations, we walk backwards through the neural network to generate deltas for each neuron. [12]. Finally, for each synapse, the output delta is multiplied by the input activation to get a gradient for the weight, which is then modified by a ratio of the percentage of this gradient. [12]. This is repeated until the network converges.

For example, let $\sigma(\mathtt{x}) = \frac{1}{1+\sigma^(-\mathtt{x})}$ be the sigmoid activation function. The derivative of the sigmoid is calculated as follows.

$$\sigma'(\mathtt{x}) = \frac{\mathtt{d}}{\mathtt{dx}}\left(\frac{1}{1+\sigma^{-\mathtt{x}}}\right) \tag{2.8}$$

$$\sigma'(\mathtt{x}) = \frac{1+\mathtt{e}^{-\mathtt{x}}\frac{\mathtt{d}}{\mathtt{dx}}(1) - 1\frac{\mathtt{d}}{\mathtt{dx}}(1+\mathtt{e}^{-\mathtt{x}})}{(1+\mathtt{e}^{-\mathtt{x}})^2} \tag{2.9}$$

$$\sigma'(\mathtt{x}) = \frac{0-(-\mathtt{e}^{-\mathtt{x}})}{(1+\mathtt{e}^{-\mathtt{x}})^2} \tag{2.10}$$

$$\sigma'(\mathtt{x}) = \frac{\mathtt{e}^{-\mathtt{x}}}{(1+\mathtt{e}^{-\mathtt{x}})^2} \tag{2.11}$$

$$\sigma'(\mathtt{x}) = \frac{(1+\mathtt{e}^{-\mathtt{x}})-1}{(1+\mathtt{e}^{-\mathtt{x}})^2} \tag{2.12}$$

$$\sigma'(\mathtt{x}) = \frac{1+\mathtt{e}^{-\mathtt{x}}}{(1+\mathtt{e}^{-\mathtt{x}})^2} - \left(\frac{1}{1+\mathtt{e}^{-\mathtt{x}}}\right)^2 \tag{2.13}$$

$$\sigma'(\mathtt{x}) = \frac{1}{1+\mathtt{e}^{-\mathtt{x}}} - \left(\frac{1}{1+\mathtt{e}^{-\mathtt{x}}}\right)^2 \tag{2.14}$$

$$\sigma'(\mathtt{x}) = \sigma(\mathtt{x}) - \sigma(\mathtt{x})^2 \tag{2.15}$$

$$\sigma'(\mathtt{x}) = \sigma(\mathtt{x})(1-\sigma(\mathtt{x})) \tag{2.16}$$

$$\tag{2.17}$$

where $\mathtt{x} \in \mathbb{R}$

The activation for a neuron is calculated by $\mathtt{O} = \sigma(\hat{\mathtt{w}} \cdot \hat{\epsilon} + \theta)$ where $\hat{\mathtt{w}}$ is the edge weight vector and $\epsilon$ is the input and $\theta$ is the bias. A bias node is simply an input to a neuron that offsets the activation threshold by $\theta$. The bias can be thought to be an input neuron with an output of 1 and a weight of $\theta$. For example, if we had 3 inputs to a neuron, $\epsilon_1, \epsilon_3, \epsilon_3$ with weights $\mathtt{w}_1, \mathtt{w}_2, \mathtt{w}_3$, then the activation would be computed with $\mathtt{O} = \sigma(\mathtt{w}_1 \cdot \epsilon_1 + \mathtt{w}_2 \cdot \epsilon_2 + \mathtt{w}_3 \cdot \epsilon_3 + \theta)$ or more succinctly $\mathtt{O} = \sigma(\hat{\mathtt{w}} \cdot \hat{\epsilon} + \theta)$. Recall that $\sigma(\mathtt{x}) = \frac{1}{1+\mathtt{e}^{-\mathtt{x}}}$ and $\sigma'(\mathtt{x}) = \sigma(\mathtt{x})(1-\sigma(\mathtt{x}))$

$x^l_j$: input to node j of layer l
$W^l_{ij}$: weight from layer l-1 node i to layer l node j
$\theta^l_j$: bias of node j of layer l
$t_j$: target value of node j of the output layer

The following notation will be used for the rest of this section.

Given a set of training points $t_j$ and output layer $O_j$ error is defined as

$$E = \frac{1}{2}\sum_{k \in K}(O_k - t_k)^2 \tag{2.18}$$

where K is every node in the output layer, and J is every node in the hidden layer.

$$\frac{\partial E}{\partial W_{jk}} = \frac{\partial}{\partial W_{jk}}\frac{1}{2}\sum_{k \in K}(O_k - t_k)^2 \tag{2.19}$$

$$= (O_k - t_k)\frac{\partial}{\partial W_{jk}}O_k \tag{2.20}$$

$O_k$ is just the activation of the neuron, which is simply $\sigma(x_k)$

$$= (O_k - t_k)\frac{\partial}{\partial W_{jk}}\sigma(x_k) \tag{2.21}$$

$$= (O_k - t_k)\sigma(x_k)(1 - \sigma(x_k))\frac{\partial}{W_{jk}}x_k \tag{2.22}$$

$$= (O_k - t_k)O_k(1 - O_k)O_j \tag{2.23}$$

because $\sigma(x_k)$ is simply the output $O_k$ and $x_k$ is simply the output $O_j$

Let $\delta_{\mathtt{k}} = (\mathtt{O_k} - \mathtt{t_k}) * \mathtt{O_k} * (\mathtt{1} - \mathtt{O_k})$ then $\frac{\partial \mathtt{E}}{\partial \mathtt{W_{jk}}} = \mathtt{O_j}\delta_{\mathtt{k}}$ which gives us the update to the edge weights for the output layer. Now let us work backwards to the hidden layer.

$$\frac{\partial \mathtt{E}}{\partial \mathtt{W_{ij}}} = \frac{\partial}{\partial \mathtt{W_{ij}}} \frac{1}{2} \sum_{\mathtt{k} \in \mathtt{K}} (\mathtt{O_k} - \mathtt{t_k})^2 \tag{2.24}$$

$$= \sum_{\mathtt{k} \in \mathtt{K}} (\mathtt{O_k} - \mathtt{t_k}) \frac{\partial}{\mathtt{W_{ij}}} \mathtt{O_k} \tag{2.25}$$

$$= \sum_{\mathtt{k} \in \mathtt{K}} (\mathtt{O_k} - \mathtt{t_k}) \sigma(\mathtt{x_k})(\mathtt{1} - \sigma(\mathtt{x_k})) \frac{\partial}{\mathtt{W_{ij}}} \mathtt{x_k} \tag{2.26}$$

We need to rewrite $\frac{\partial}{\mathtt{W_{ij}}} \mathtt{x_k}$ to be in terms of $\mathtt{O_j}$ using the chain rule.

$$= \sum_{k \in K} (O_k - t_k)\sigma(x_k)(1 - \sigma(x_k)) \frac{\partial x_k}{O_j} \frac{\partial O_j}{\partial W_{ij}} \tag{2.27}$$

$$= \sum_{k \in K} (O_k - t_k)\sigma(x_k)(1 - \sigma(x_k)) w_{jk} \frac{\partial O_j}{\partial W_{ij}} \tag{2.28}$$

$$= \frac{\partial O_j}{\partial W_{jk}} \sum_{k \in K} (O_k - t_k) O_k (1 - O_k) w_{jk} \tag{2.29}$$

$$= \frac{\partial O_j}{\partial W_{jk}} \sum_{k \in K} \delta_k w_{jk} \tag{2.30}$$

$$= \frac{\partial \sigma(x_j)}{\partial W_{jk}} \sum_{k \in K} \delta_k w_{jk} \tag{2.31}$$

$$= O_j(1 - O_j) \frac{\partial x_j}{\partial W_{ij}} \sum_{k \in K} \delta_k w_{jk} \tag{2.32}$$

$$= O_j(1 - O_j) O_i \sum_{k \in K} \delta_k w_{jk} \tag{2.33}$$

$$\tag{2.34}$$

Then

$$\delta_j = O_j(1 - O_j) \sum_{k \in K} \delta_k w_{jk} \tag{2.35}$$

$$\frac{\partial E}{\partial W_{jk}} = O_i \delta_j \tag{2.36}$$

This means that the back propogation algorithm is as follows.

1. Run the forward step

2. For each output node compute $\delta_k$

3. For each hidden node calculate $\delta_j$

4. Update the weights and biases with $\Delta w = -\nu \delta_l O_{l-1}$ and $\Delta\theta = -\nu\delta_l$ and apply $w + \Delta w \rightarrow w$ and $\theta + \Delta\theta \rightarrow \theta$ where $\nu$ is a hyperparemeter representing the learning rate of the network.

### 2.3.4 Forward Activation in an RNN

Forward activation in an RNN works similarly to a FFNN, except you have one extra parameter. Let $\hat{i}$ be the inputs, $\hat{w_i}$ be the edge weights, $\hat{w_y}$ be the edge weights from the input vector from the previous activation, and $\hat{y}$ be the input from the previous activation. The forward activation is then calculated by $F(\hat{i} \cdot \hat{w_i} + \hat{y} \cdot \hat{w_y})$. Likewise, the activation function $F(x)$ is typically a smooth differentiable function. For the backward step it is best if you think of an RNN as a FFNN with time; you can unroll the copies of the network that happen over time. With this, back propagation through time works just like normal backpropagation, albeit with some problems. In a feed forward neural network we keep the parameters stable by using a squashing function such as sigmoid to keep the activation $0 <= x <= 1$ where $x \in \mathbb{R}$ This does not work in the backwards direction. For backpropagation, we are multiplying all of the gradients in each layer together meaning the further backwards you go, the less impact the error calculation has on the network. This is not a problem with RNNs, it is a problem with all neural networks that use gradient descent. A deep enough feed forward neural network will also see this problem. RNNs just see it exponentially quicker. One common technique for dealing with this is to use a modified gradient descent called truncated back propagation through time which essentially cuts off the back propagation after N time steps, having the negative side effect of not letting us make temporal connections that are far apart, but the positive side effect of making the network trainable. It is worth noting these hyperparameters are either arbitrary or tunable to a certain application.

### 2.3.5 Long Short-Term Memory

RNNs have been described as fundamentally difficult to train due to their iterative nature; error signals flowing backwards would either compound or disappear completely. [22] Long Short-term Memory (LSTM) learns to bridge these time intervals. LSTM was developed in order to solve this vanishing gradient problem [22]. This problem was solved by creating a memory unit. One input is called the input block, which is conditionally multiplied by an input gate. The input gate is there to conditionally determine when the value is allowed to propagate through the network. See Figure 2.5. The forget gate, when zero, zeros the memory block, while the output gate determines if the memory block should output what is in memory. The SUM node feeds back on itself, so that the memory block can remember the contents of memory for an arbitrary length of time. [23].

### 2.3.6 Conclusion

Machine learning techniques, as shown, lead to remarkable advances towards building more intelligent systems. The recent advent and popularity of software defined networking has created an opportunity to advance the state of the art by using computational intelligence to solve remarkably difficult problems in the world of computer networking. All of supervised, unsupervised, and reinforcement learning have applications in computer networking. Supervised learning algorithms require pre-labeled training data. Unsupervised learning algorithms have been successfully used in computer networking in order to automatically detect new attack vectors in the context of an intrusion detection system (IDS) for example [24].

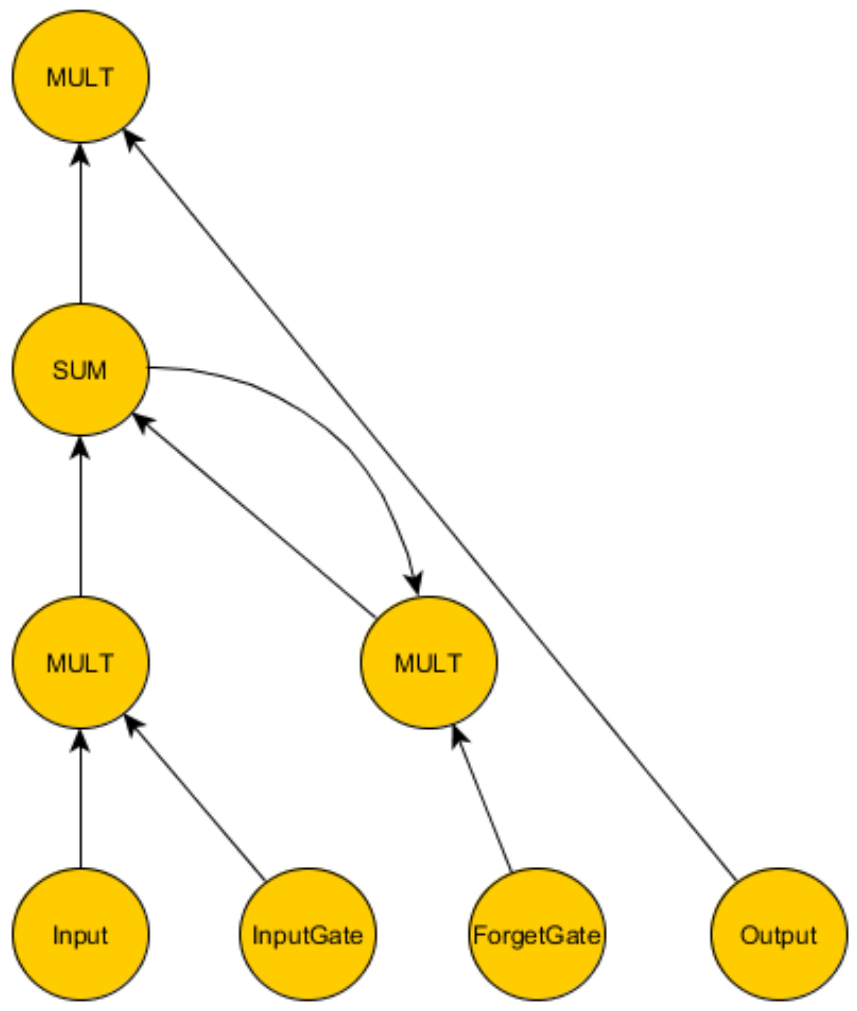

**Figure 2.5**: Long Short Term Memory

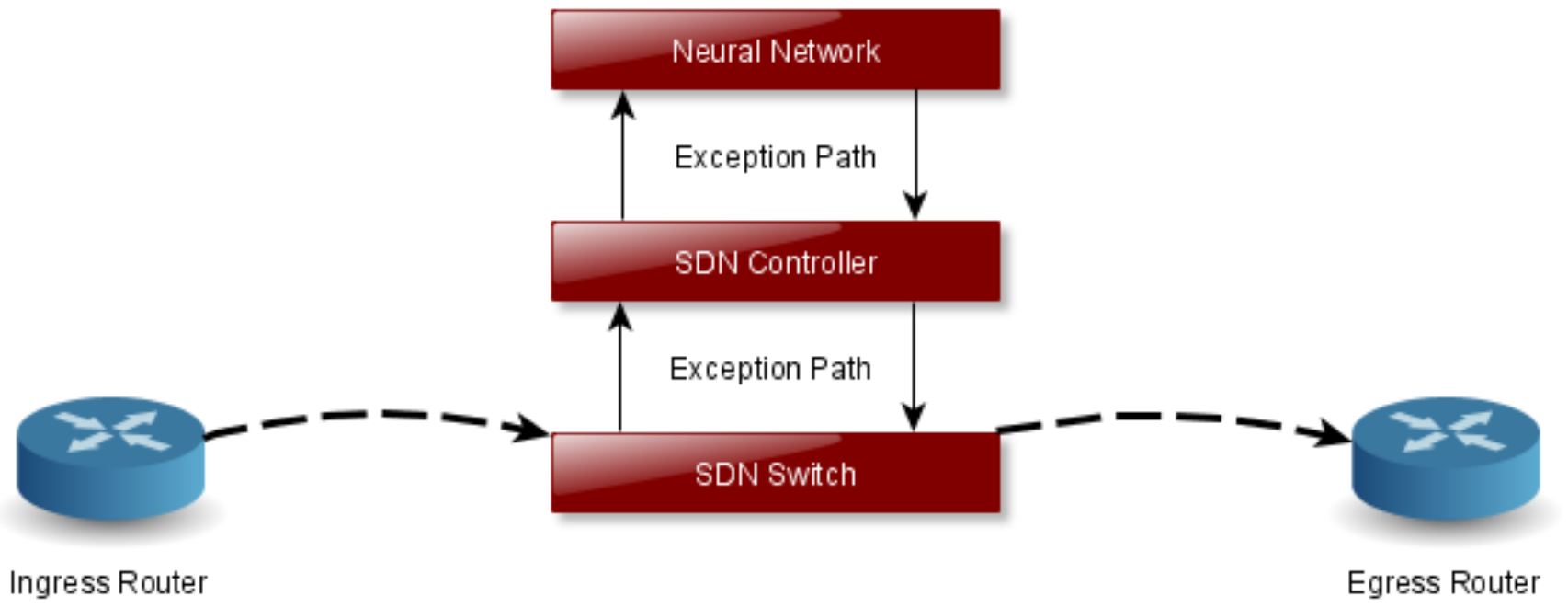

**Figure 2.6**: Overview

Consider the example in Figure 2.6. If the ingress flow table does not have an entry for the ingressing packet, it will forward the packet up the exception path to the SDN controller. The SDN controller will then need to decide what will happen to this frame and push a flow entry down. When the flow expires, the flow stats will be sent back up the exception path for the SDN controller. This is where learning can take place. In the forward propagation step, the neural network made a prediction about the size of a flow; how many frames were passed over the life of the flow. When the flow expires we get the actual statistics for this flow letting us update our model. The SDN controller made a decision about what should happen to this flow, and based on the flow statistics, we can provide feedback to the neural network for back propagation. In this model, the controller would be the agent, the neural network would have the precepts, and the action would be decided by the SDN controller. This is the technique we utilized in this thesis. The neural network makes prediction about the utility of a flow, and adjusts the model to compensate for error when it learns more (upon flow expiration). In the next section we will describe the architecture for achieving this.

# CHAPTER 3

# THE ROX OPENFLOW CONTROLLER

In the previous chapter we introduced OpenFlow and neural networks. This chapter ties these two concepts into a concrete software design. The ROX OpenFlow controller will serve as the control plane for a neural network, while Open vSwitch will be the forwarding plane. Throughout this chapter we will describe the architecture and run scale tests which will serve as the control for future experimentation.

## 3.1 OpenFlow Controller Microservice

From a high level, as seen in Figure 3.1, the IPV4 Traffic is sent to one of the two ports on Open vSwitch (OVS). OVS starts traversing the flow tables (see Section 2.1.3 for details). If there is a flow match, the specified action is taken and the packet is forwarded out the designated egress port. For instance, consider Figure 3.2. In this example, any packet with a ip source address of 192.168.2.46, and ip dest address of 192.168.1.46, with L4 protocol 253 will have the L2 source address set to 52:00:00:00:00:aa, and the L2 dest address set to 52:00:00:00:00:01, and be transmitted out port 2 if and only if no other flows have a priority greater than 10000. An example of an Exception Entry is shown in Figure 3.3. If there are no Flow Table matches, and there are entries in the flow table, the switch will drop the packet. If the Exception Entry is matched, the frame will be wrapped in a PACKET_IN openflow header, and forwarded to ROX. See Figure 3.4 for an example.

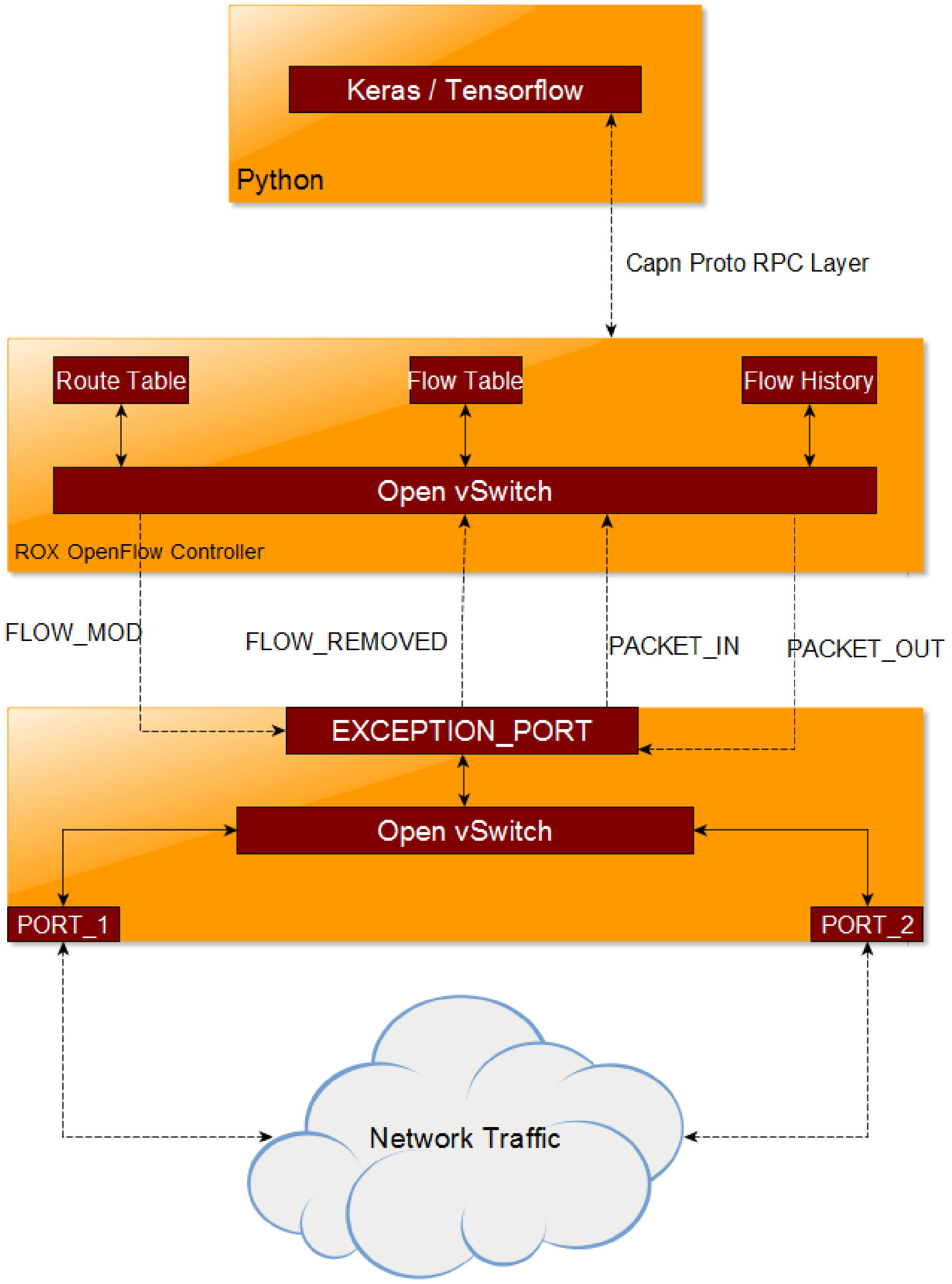

**Figure 3.1**: Overview of ROX

```
cookie=0x2dede2b6, duration=10.481s, table=0, n_packets=27, n_bytes=1620, idle_timeout=10, hard_timeout=20,
        send_flow_rem priority=10000,
        ip,nw_src=192.168.2.46,nw_dst=192.168.1.46,nw_proto=253,nw_tos=0
        actions=set_field:52:00:00:00:00:01->eth_dst,set_field:52:00:00:00:00:aa->eth_src,output:2
```

**Figure 3.2**: Flow Table Example

```
cookie=0x0, duration=14.172s, table=0, n_packets=0, n_bytes=0, priority=1,in_port=2 actions=CONTROLLER:65535
```

**Figure 3.3**: Exception Entry Example

```
▸ Frame 163973: 168 bytes on wire (1344 bits), 168 bytes captured (1344 bits) on interface 0
▸ Ethernet II, Src: 00:00:00_00:00:00 (00:00:00:00:00:00), Dst: 00:00:00_00:00:00 (00:00:00:00:00:00)
▸ Internet Protocol Version 4, Src: 127.0.0.1, Dst: 127.0.0.1
▸ Transmission Control Protocol, Src Port: 37650 (37650), Dst Port: 6653 (6653), Seq: 9857, Ack: 15569, Len: 102
▾ OpenFlow 1.4
    Version: 1.4 (0x05)
    Type: OFPT_PACKET_IN (10)
    Length: 102
    Transaction ID: 0
    Buffer ID: OFP_NO_BUFFER (0xffffffff)
    Total length: 60
    Reason: OFPR_APPLY_ACTION (1)
    Table ID: 0
    Cookie: 0x0000000000000000
  ▸ Match
    Pad: 0000
  ▾ Data
    ▸ Ethernet II, Src: RealtekU_dc:53:01 (52:54:00:dc:53:01), Dst: RealtekU_dc:52:01 (52:54:00:dc:52:01)
    ▸ Internet Protocol Version 4, Src: 192.168.2.97, Dst: 192.168.1.97
    ▸ Data (26 bytes)
```

**Figure 3.4**: PACKET_IN Message

Once the frame reaches the controller, it will parse the type field from the OpenFlow Header, and send it to the relevant subsystem. For a PACKET_IN Message, the Data Segment of the OpenFlow Message is send to the route table for a next hop address. The ARP table is checked to resolve the L2 address for the next hop.

This information is then sent to the Flow Table, where an FLOW_MOD Message is generated. A copy of this information is stored, along with a timestamp, in a cache for future processing, and the message itself is sent back to the flow switch for processing. The flow switch will then create the appropriate entry in the flow table. Once the expire time has expired, the switch will then send a FLOW_REMOVED message which will contain the cookie, which is a controller chosen index for the flow, and various packet statistics like packet count, duration, and byte count; it is the packet count that the neural network will be predicting. The packet count is the number of frames that were forwarded over the lifetime of the flow entry on Open vSwitch. It is this metric that is used to determine how important a flow is; the more frames that matched a flow entry, the more important the flow. When the controller receives the FLOW_REMOVED message it sends it to the flow table so it can remove the entry in the local flow cache, and update the flow stats entry with the flow statistics, finally pushing these stats onto a linked list for later use by the NN for back propagation. To clarify, we are concerned with the packet count, but another useful item in the flow stats entry would be Byte count. See Sections 2.1.4 and 2.1.6 for more information about these message types.

## 3.2 Neural Network Microservice

ROX is a microservices based architecture with communication happening across an RPC layer utilizing Cap'n Proto (described in the next chapter). See Figure 3.1. On new flows an RPC call will be passed from the openflow controller to a python application for classification. A message will be returned with the importance of the flow. Once a flow expires the packet count will be transmitted across this RPC layer for training purposes. We choose this architecture in order to allow us to separate the neural network and OpenFlow controller into two separate pieces of software that function mostly independently.

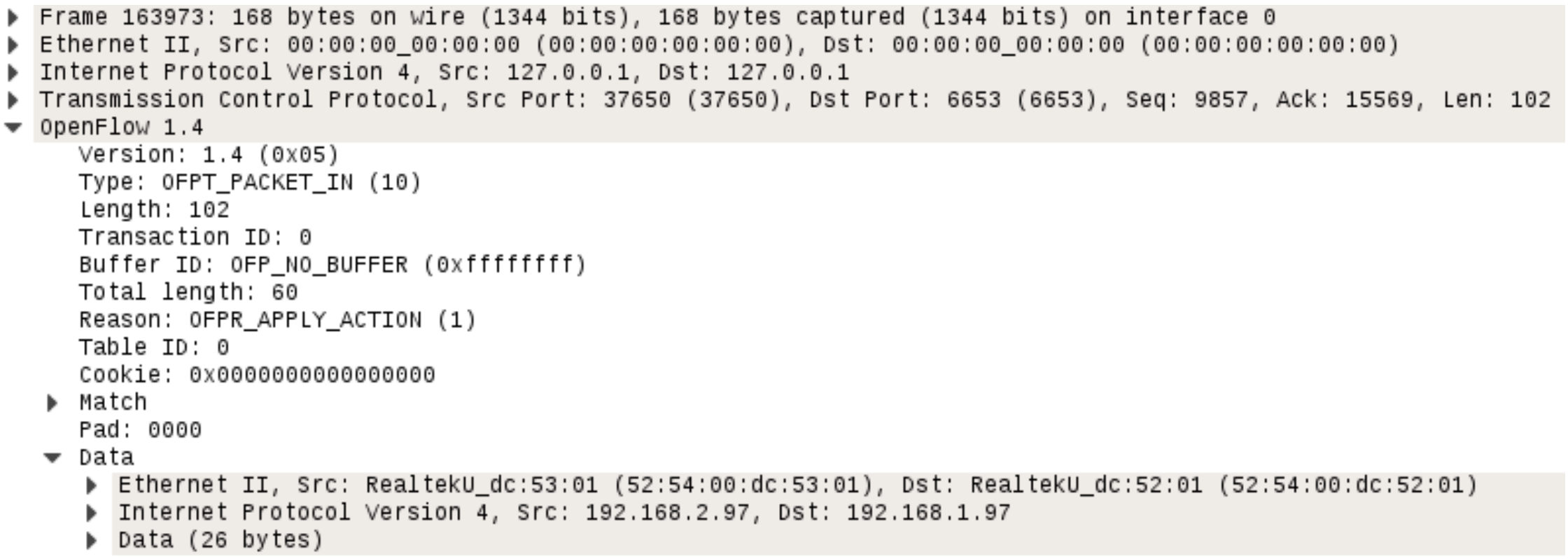

**Figure 3.5**: FLOW_MOD Message

# CHAPTER 4

# NEURAL NETWORK FOR FLOW UTILITY PREDICTION

## 4.1 Introduction

In order to determine the destination of a packet in a packet based routing network, one must inspect the IP header of the packet, once that is done, a lookup is performed in the routing table in order to determine the next hop L2 address based on the destination payload in the IP header. Finally, the packet is forwarded out the correct interface. The problem with this approach is that it is repetitive, as the next-hop address is going to be the same for a given flow. Flow based routing instead builds a hash of the 5-tuple (dest address, source address, dest port, source port, and protocol) so that the same action is applied to each packet in the flow, removing route-table lookup from the control-plane. There is a cost involved in creating and maintaining the flow map. You have to keep track of the age, and hit counts for the flows, which uses resources on a re-occurring basis. This means that if a flow only has one packet, you are using more resources by processing the packet in a hardware flow map than you would be just letting it go to the CPU. An example of this is DNS flows. Typically a DNS flow will have a couple of packets, and then never be used again,

**Table 4.1**: Components and Libraries

| Component | Version |
|---|---|
| Capnp (Cap'n Proto) | 0.5.3 |
| Keras | 1.2.1 |
| Tensorflow | 0.12.1 |
| C++ Standard | C++11 |
| GCC | 6 |
| Python | 2.7.13 |

using space in the hardware flow cache that could be used for more high-throughput flows.

The problem is two fold. First, we need to be able to predict when a flow will have only one hit, and never push those flows into the hardware flow cache. Second, we need a way to prioritize high-throughput flows, and make sure those flows go into the flow cache ahead of low-throughput flows. In this section we will demonstrate a flow simulator, and a neural network approach to prioritizing high-throughput flows into the flow cache, and demonstrate the simulated performance gain.

## 4.2 Test System

All tests were run on a dual-cpu hexacore xeon system with 128GB of memory running Debian (Jessie) Linux. The RPC layer that ties the neural network to the ROX OpenFlow controller was done using Cap'n Proto version 0.5.3. Cap'n proto is a capability-based RPC system similar to Protocol Buffers. The neural network was coded using Keras version 1.2.1 and Tensorflow version 0.12.1. The ROX Openflow controller was written in C++11 using GCC 6 as the compiler. A link to the source code is available in the appendix. All datasets can be found on github as mentioned in the appendix. The data used in these tests were packet header captures on a live production network. One week's worth of data was captured in order to get a full range of traffic patterns. The captures are stored on Github, as seen in the appendix, for reproducibility purposes.

”Tensorflow is an open source library for numerical computation using data flow graphs. Nodes in the graph represent mathematical operations, while the graph edges represent the multidimensional data arrays (tensors) communicated between them” [25] Tensorflow is a backend for Keras which are both used in this thesis. ”Keras is a high-level neural networks API written in python and capable of running on top of either TensorFlow or Theano.” [26]

## 4.3 Feed Forward Neural Network

### 4.3.1 Test Setup and Methodology

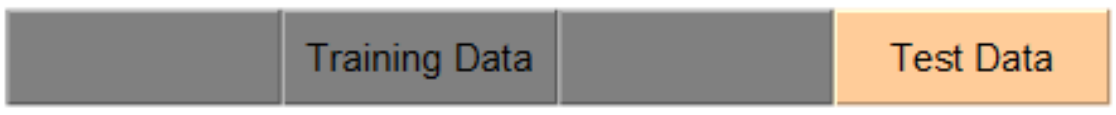

**Figure 4.1**: Data Layout

This test will send recorded data from a live network to the NN through the Openflow controller. The recorded data includes the flow header and packet counts. The results are measured by simulating a limited flow table. We are assuming we have half as much space in the flow table as we do flows and measure the number of packets that would be in the preferred half based on the NN prediction. The goal of the neural network is to sort the largest flows into the top half of the flow table, and the smallest flows into the bottom half of the flow table, as per research questions 1 and 2 in chapter 1 section 1.1.3. If the algorithm is successful we would expect to see both accurate assortment into the top and bottom half of the flow table and a speedup over a first come first serve assortment. In this context speedup denotes an increase in packet throughput over each 3247 flow interval and is measured by determining the number of packets that are sorted into the top half of the flow table using the neural network and dividing that by the number of packets that made it into the top half of the flow table without using the neural network. The neural network used for this is a feed forward neural network with 16 input neurons, 3 hidden layers with 50 neurons a each, and 5 output neurons that are one-hot coded. A neural network essentially gives you a probability that a given node should be activated. In a one-hot encoding, only one node can be active at any given time. In Figure 4.2, the node with value 0.76 will be activated. We use such a coding scheme for the neural network in this project. The output neuron corresponds to the correct priority, where priority is such a one hot code. Dropout layers were added with a dropout rate of 0.2 between each hidden layer to help prevent over-fitting. Dropout layers add a random chance that neurons will misfire forcing the network to build multiple internal models. The loss function in the neural network was implemented as categorical cross entropy. Categorical cross entropy is defined as $\mathtt{H}(\mathtt{p},\mathtt{q}) = -\sum_{\mathtt{x}} \mathtt{p}(\mathtt{x})\mathtt{log}(\mathtt{q}(\mathtt{x}))$, where p is a tensor in $\mathbb{Z}^2$ and q is a tensor in $\mathbb{R}^2$. "This measures the average number of bits needed to identify an event from a set of possibilities." [27] This algorithm makes sense as we are trying to classify into N categories, where N is the number of priorities. The learning algorithm for the neural network was the Adadelta learning algorithm, as mentioned by Zeiler [28] in order to update the gradients. The input and hidden layers are using a rectified linear activation function as described by Nair et al [29], while the output neurons are using the sigmoid activation function as previously described in section 2.3.1. We then show both speedup over time and classification accuracy. This is the same definition of speedup used earlier. Speedup was demonstrated with live data captured on a live network, while the accuracy replayed the saved flow header information retraining the network on 75% of the data set, and testing it on the remaining 25%.

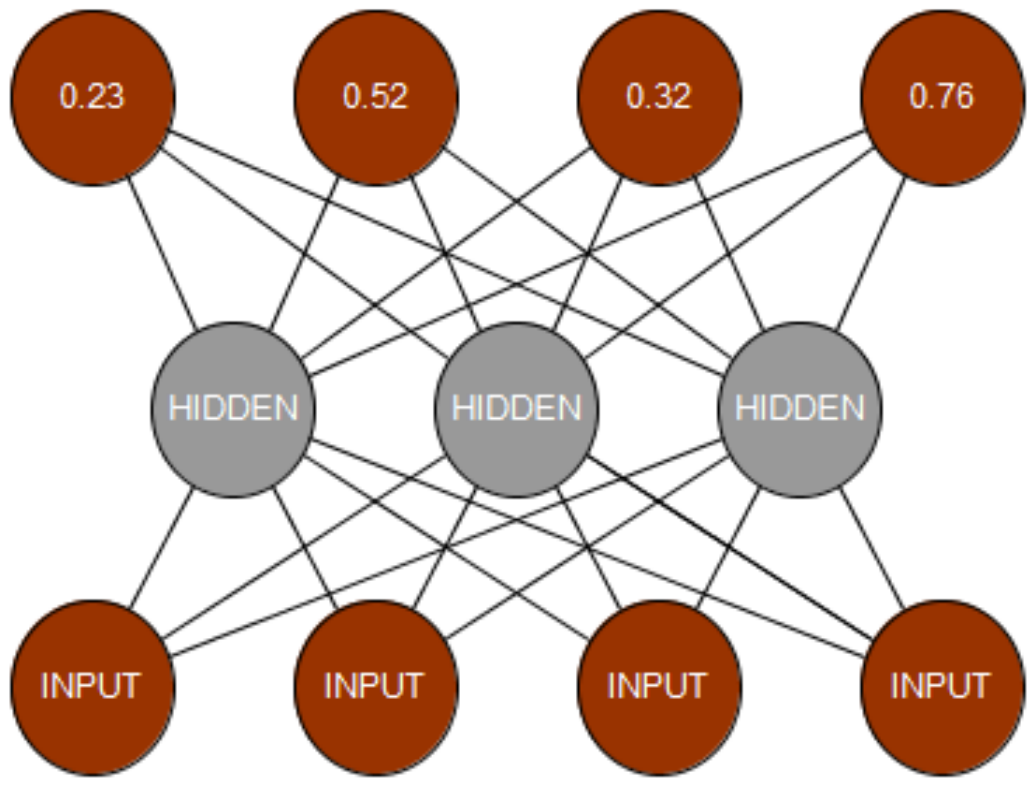

**Figure 4.2**: One-hot encoding

### 4.3.2 Classification Accuracy

For this test we iterated through the data set in slices of 3247 training examples at a time with a test set containing 1083 examples that are never shown to the neural network. Each slice of 3247 examples was trained for 5 epochs. The FFNN Classification Accuracy is defined as the number of correct guesses vs the total number of guesses. After 1 iteration of 5 epochs the network had reached an accuracy of 74.9 percent with an standard deviation of 0.02 across 30 trials. The neural network struggles in the center, but that is likely due to changing data patterns, or it is data that does not have a lot of variance. It is worth noting the accuracy remains above 60 percent most of the time, and eventually re-converges at above 80 percent. At iteration 1, we do get most of the data residing in the top half of the flow table, but with a high standard deviation, as shown in the error bars. As the network learns, the standard deviation shrinks, and it widens again in this section in the middle. The neural network eventually returns to stay above 80 percent until the end. The most useful chart for demonstrating the perfmance of the neural network is the speedup chart. Speedup is defined as the number of packets that made it into half the flow table with the neural network vs how many would make it into the neural network with first come first serve. In this chart the speedup was typically at 2, with some spikes due to first come first serve getting extremely unlucky.

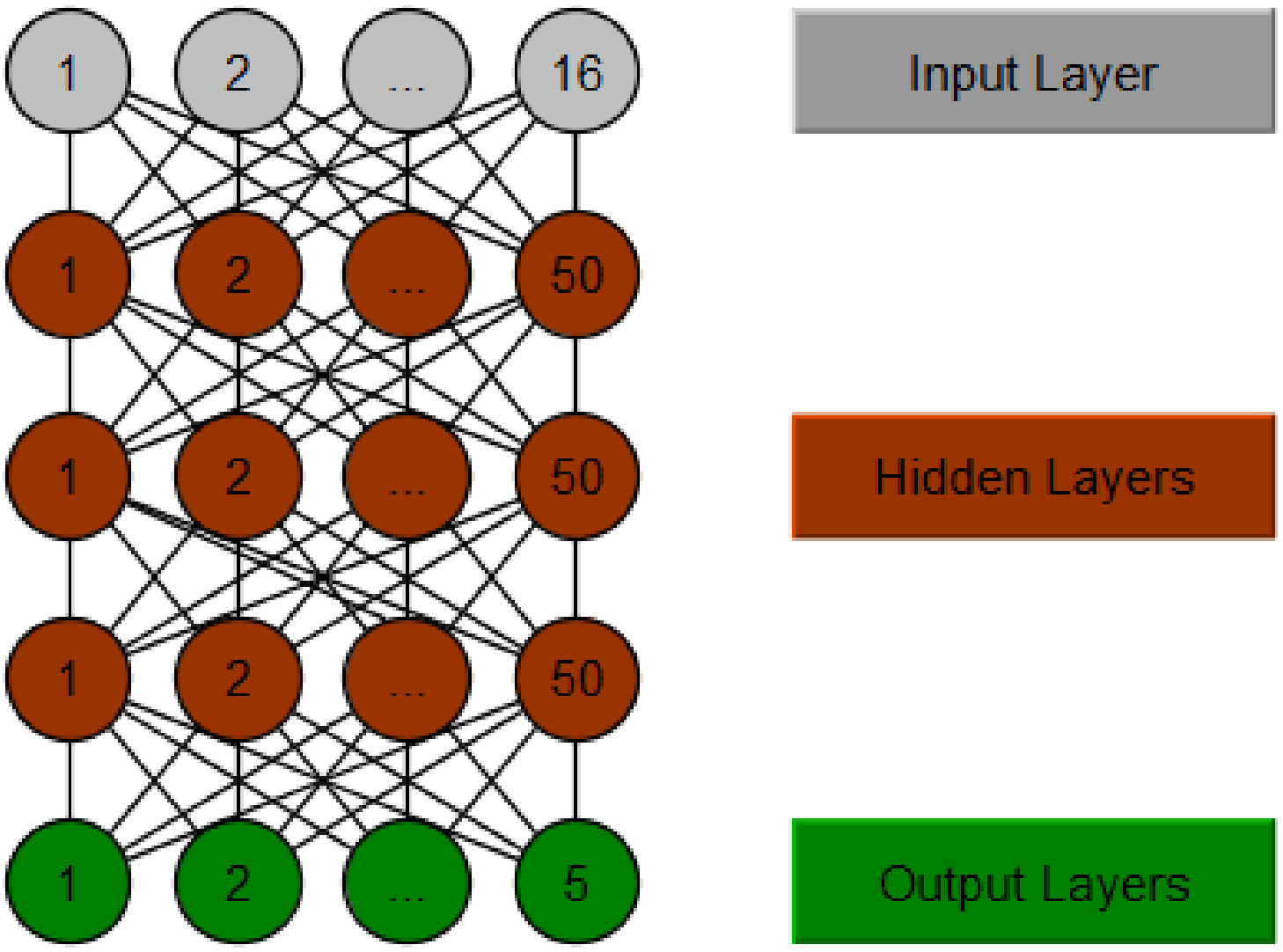

**Figure 4.3**: Structure of the Feed Forward Neural Network

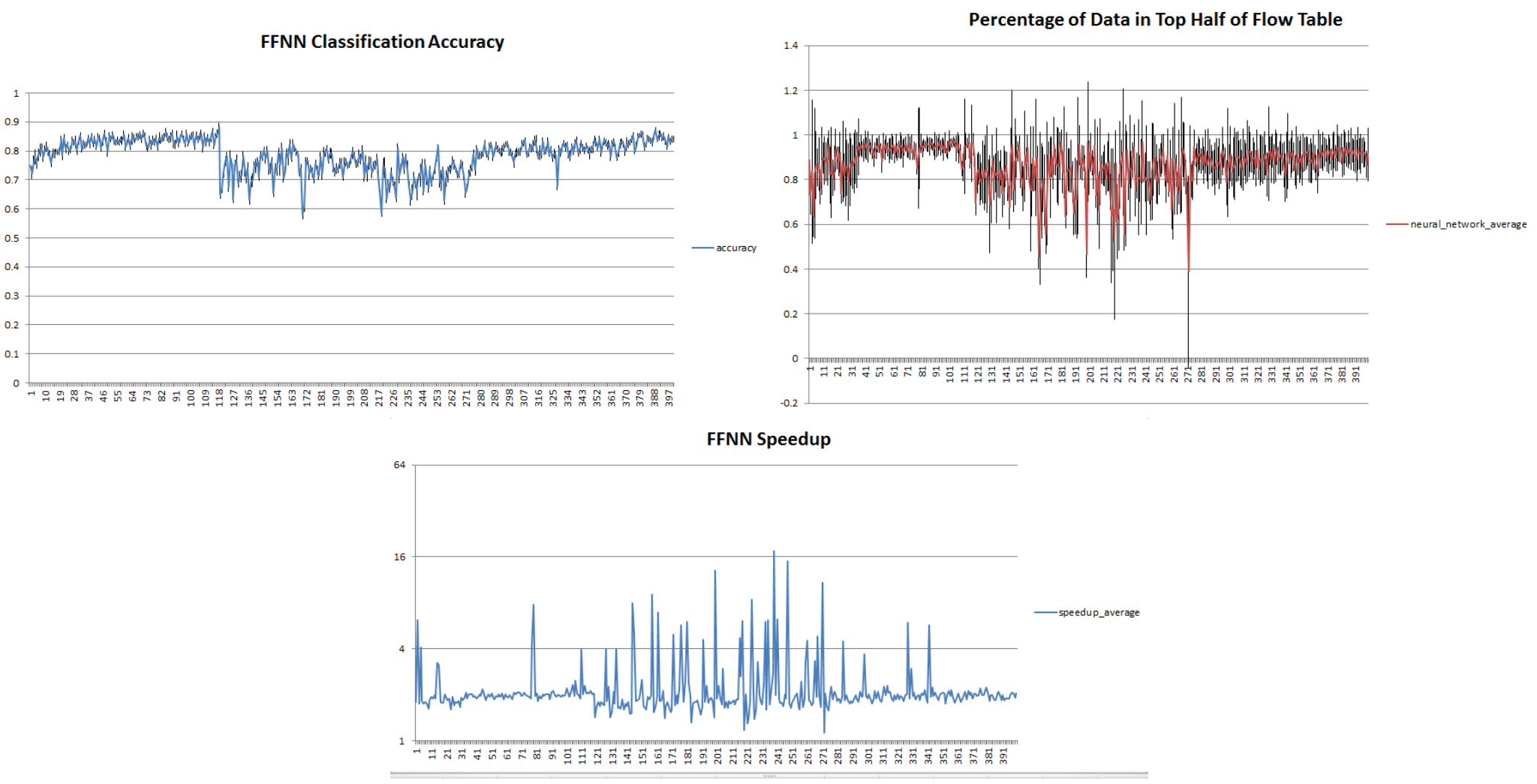

**Figure 4.4**: Feed Forward Neural Network Results

This classifier, as shown in the the data included at the github repository in the appendix, and Figure 4.6 performs well in a feed forward network and demonstrates remarkable performance gains. As noted in previous sections, this is especially useful in systems that use hardware LUTS (Lookup Tables) for the flow tables with extremely limited space. Most of the flows are small flows, as also shown in the data included at the github repository in the appendix, which implies they would be better handled in software due to the fact that you have to process the packets at least once in software, and for small once off flows such as DNS never again.

## 4.4 Long Short Term Memory Neural Network

### 4.4.1 Test Setup and Methodology

This test will use the same dataset collected in the previous test through a LSTM neural network in order to compare the performance of this network with the FFNN. If there is context sensitive bits of information in the flow header, we would expect to see superior performance. If there are no context sensitive bits, we would see marginal improvements, none at all, or reduced performance due to over-fitting.

### 4.4.2 Classification Accuracy

This test is identical to the test shown in the previous section, just with a long short term memory network instead of a feed forward network. The results are also essentially identical. The reason for this is that our network is only looking at first packets. First packets are the packets that initiate a flow entry in the flow table. Since we are only looking at first packets, there is little context that an LSTM can use to increase accuracy. It is worth noting that the training time on the LSTM network was approximately 10 times longer, making this a poor choice.

## 4.5 OpenFlow Controller Revisited

One of the stated goals of this research is to determine if we could utilize this approach to build a more intelligent open-flow controller. To demonstrate this we created 5 importance levels with 1 being low importance and 5 being high importance. Importance is the number of flows divided into 5 categories. In terms of the neural network, this maps to one of the output neurons on the neural network. We are just calculating this with perfect knowledge instead of using the neural network; we know already how many packets will be in a flow. We took one snapshot of the flow table in Open vSwitch using the command ovs-ofctl dump-flows and showed 27 importance 5 flows, 11 importance 4 flows, 310 importance 2 flows and 626 importance 1 flows. Recall that importance is the name of the field that Open vSwitch uses to determine eviction preference; low importance flows are evicted from the flow table before high importance flows. This data demonstrates that only a small number of flows are really critical to be handled in hardware.

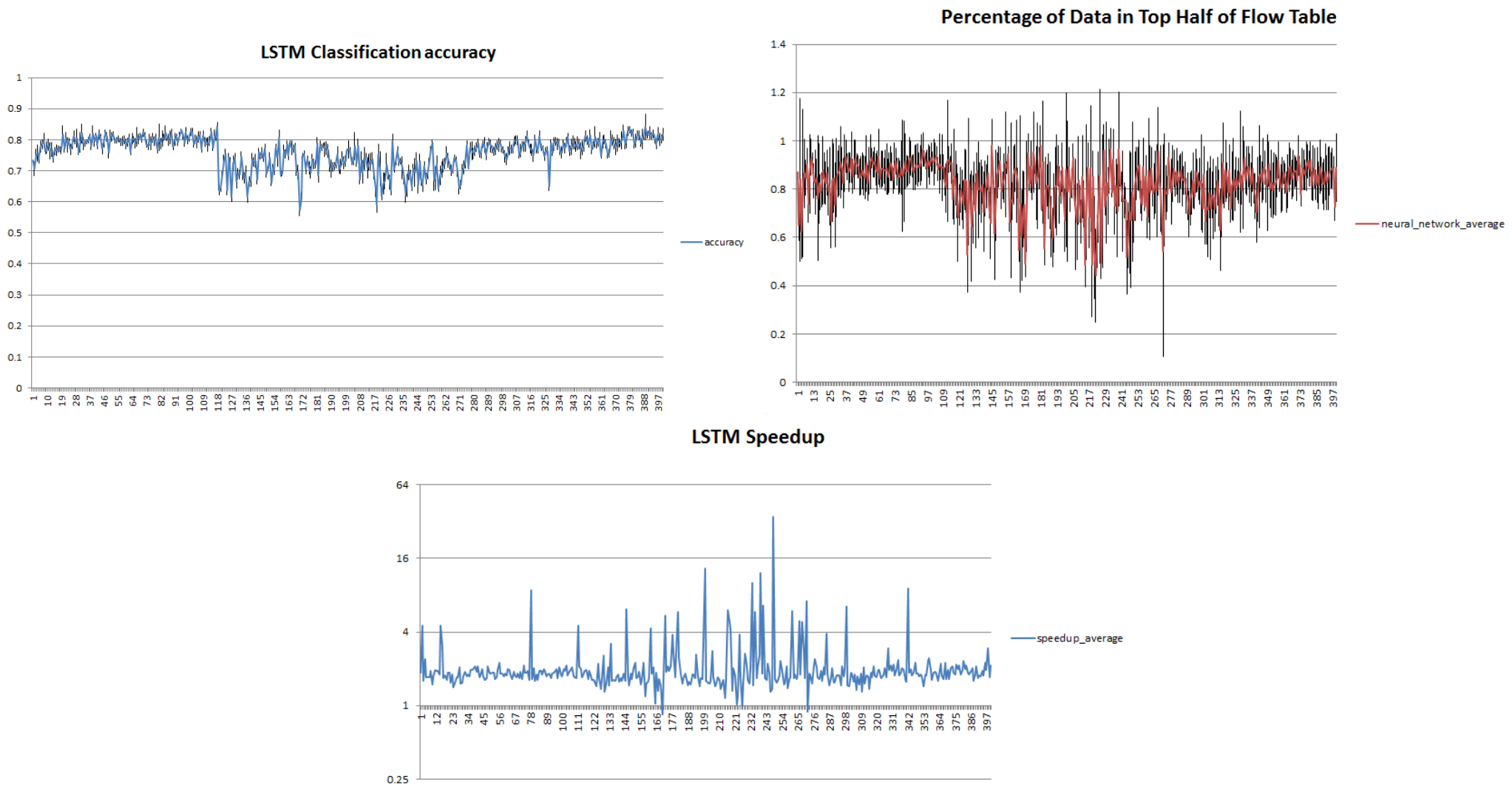

**Figure 4.5**: LSTM Neural Network Results

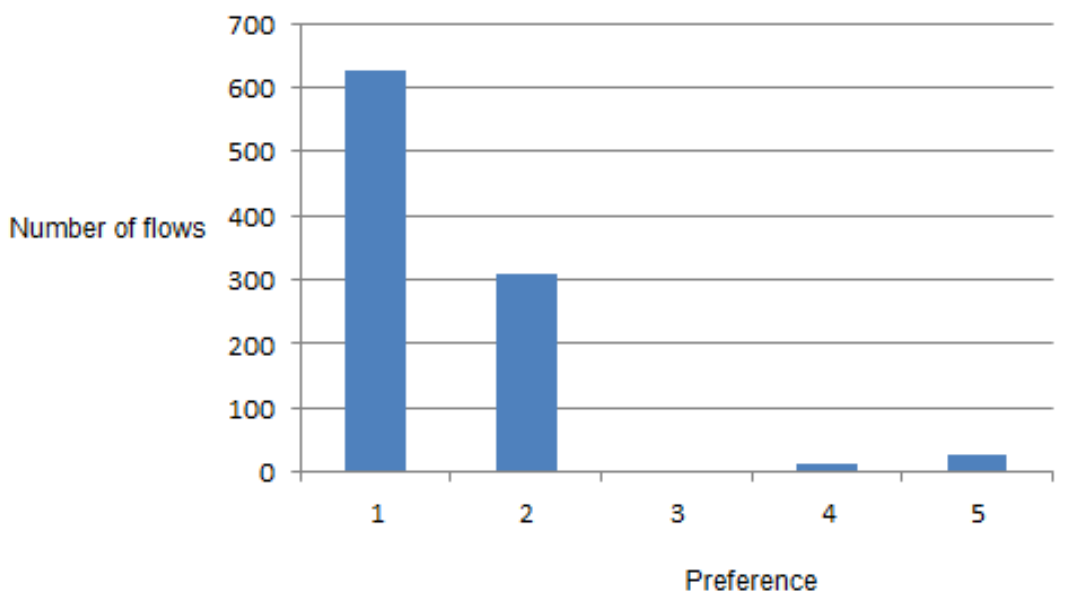

**Figure 4.6**: Flow distribution in the OpenFlow Controller

# CHAPTER 5

# CONCLUSIONS

In this thesis we built a prototype for a Neural Network driven Openflow controller. We examined one particular use case for such a system where we have a software data path (slow path) and a hardware data path (fast path). In such a system the hardware data path is fast, but has limited resources compared to the software data path. Not only that, packets have to be processed by the software path before the flow entries can be pushed down to the hardware TCAMs. Because of this, flows that do not contain lots of packets waste hardware resources, as they will be marginally used, or not reused at all; DNS flows for example will likely never be reused. To solve this problem we demonstrated a system which can predict how many packets a flow will contain over the lifetime of the flow in order to answer questions 1 and 2 in section 1.1.3. We demonstrated that this system can classify flows with 87 percent accuracy in a FFNN with a speedup of between 1.7 and 2.0 over random data. We also determined that LSTM had identical performance to the FFNN given we were only working with packet headers–no payloads–and only first packets.

The research questions from the introduction are as follows:

1. Can a neural network match patterns in flow data in real time and be used to optimize the use of resources?
   - Yes, as demonstrated from the data in the previous sections.
2. Can a neural network predict future resource utilization based on current inputs?
   - Yes, as demonstrated from the data in the previous section.
3. Can these techniques be applied to dynamically and preemptively route around predicted points of congestion?
   - Yes, the ability to predict how large a flow is implies that you can determine a flows impact on resources such as load balancing, however we would need to use the entire packet including the payload to have a shot at taking into account context for more advanced analytics.

This answers the first and second stated goals of this research. The first goal was to see if we could pattern match in real time showing resource improvements. The answer to this is yes, as demonstrated. The second question was to determine if we could predict resource utilization ahead of time (before we actually got a flow). The answer to this is also yes, as previously demonstrated. The network was predicting utility and sorting it at the time the first packet was seen for flow creation. The LSTM, also worked, but it showed no performance gains while taking longer to train. The third research question was to determine if these techniques could be applied to dynamically route around predict points of congestion. The answer to question 3 is yes, the ability to predict how large a flow is implies that you can determine a flows impact on resources such as load balancing, however we would need to use the entire packet including the payload to have a shot at taking into account context for more advanced analytics.

Since beginning this research, there has been work towards applying AI to SDN [30, 31] including one that was investigating a neural networks ability to predict performance in an SDN network [31]. All of this research has occured within the last 8 months.

There is a lot of of opportunity for future work in this area. One idea is to try to use a more complex neural network to predict which path should be taken through a network segment. Another idea is to look into using convolutional networks to classify data by type for DPI purposes; automatic annotation of packets. Such systems would be of great utility for security researchers and ISPs alike. You could utilize them for an Intrusion Detection System (IDS) or to pick up on data types that could be outside the bounds of a SLA (such as illicit downloads).

**APPENDICES**

# APPENDIX A

## SOURCE CODE AND DATASETS

All source code and the relevent datasets are posted at https://github.com/arnolmi/ROX. An overview of each directory and the files contained is below. Please contact me at mka0007@uah.edu for any additional information or with any questions.

**Contents of Github:**

- NN: Contains the code for the Neural network microservice, the simulations, and the C++ bindings to communicate with the openflow controller.
- Network: Basic Network headers such as IpAddress, MacAddress, and ArpTable.
- OpenFlow: Contains the Flow Table which also builds the messages that are sent to configure the switch.
- OpenFlow/Messages: Contains decoders and encoders for all of the openflow messages used in this program.
- System: Contains some general System stuff, like hashing algorithms, the Lookup Trie for the route table, and others.